\documentstyle[aps,prl]{revtex}
\draft

\begin{document}
\author{Xiang-Song Chen, Di Qing, and Fan Wang}
\address{Department of Physics and Center for Theoretical Physics, 
Nanjing University, Nanjing 210093, China}
\title{How does baryon magnetic moment relate to its spin structure}
\date{\today }
\maketitle

\begin{abstract}
A model independent, field theoretical relation between baryon magnetic
moment and its spin structure is derived and it is explained why constituent
quark model is a good approximation in describing baryon magnetic moments.
\end{abstract}


Baryon magnetic moment is closely related to its spin structure. In
nonrelativistic quark model, ground state baryons are usually assumed to be
in pure $s$-wave and baryon magnetic moment is solely due to quark spin
contribution. By assuming constituent quark behaving as a Dirac quark, i.e.,
keeping the simple relation between quark spin $\vec \sigma$ and magnetic
moment $\vec \mu=\frac Q{2m} \vec \sigma$, then adopting only two parameters
of constituent quark masses $m_u$=$m_d$=330MeV, $m_s$=560MeV, the SU(6)
quark model explains all measured baryon magnetic moments better than 20\%.
This has been a cornerstone of constituent quark model but never has a direct 
QCD
explanation. Since 1988, polarized lepton-nucleon deep inelastic scattering
(DIS) measurements have challenged our understanding of nucleon spin
structure obtained from constituent quark model. Quark spin seems only
contributes a small amount ($\leq$30\%) of nucleon spin \cite{1}. Thus it
becomes even harder to explain the success of constituent quark model in
predicting baryon magnetic moments. And we naturally have to find out how we
can understand baryon magnetic moments with the newly discovered nucleon
spin structure.

In previous studies, some authors \cite{2,3} found that baryon magnetic
moments can be fitted with the spin structure discovered in polarized DIS as
well as or even better than in the naive quark model. In this paper, we will
start from relativistic quantum field theory and derive a complete, model
independent relation between baryon magnetic moment and its spin content,
then explain from this field theoretical relation why constituent quark
model is a good approximation at the problem of baryon magnetic moments.

We are acquainted with the relation between magnetic moment and spin,
orbital angular momentum in nonrelativistic quantum mechanics: 
\begin{equation}
\vec \mu =\frac Qm\vec S+\frac Q{2m}\vec L .  \label{1}
\end{equation}
In relativistic quantum field theory, baryon magnetic moment is defined
through its interacting with a static external magnetic field $\vec B$ (with
four vector $\vec A=\frac 12\vec B\times \vec x$, $A^0=0$): 
\begin{eqnarray}
H_I &=& \left\langle B\left| \sum_q-iQ_q\int d^3x\bar \psi _q\gamma ^\mu \psi
_qA_\mu \right| B\right\rangle \nonumber \\
&\equiv &-\left\langle B\left| \vec \mu \right|
B\right\rangle \cdot \vec B,  \label{2}
\end{eqnarray}
where color indices are implicit, $\left| B\right\rangle $ denotes the baryon
state, and the magnetic moment operator can be straightforwardly shown to be 
\begin{equation}
\vec \mu =\sum_q\vec \mu _q=\sum_q\frac{Q_q}2\int d^3x\psi _q^{\dagger }\vec 
x\times \vec \alpha \psi _q.  \label{3}
\end{equation}

To find the relation between baryon magnetic moment and its spin structure,
we expand $\psi (x)$ in terms of Dirac spinors. We must keep in mind that we
are now dealing with quarks interacting with gluons, therefore we can not
expand $\psi (x)$ in a free manner, but we can still expand it at a given
time (say $t$=0) as: 
\begin{equation}
\psi _q\left( x\right) =\frac 1{\left( \sqrt{2\pi }\right) ^3}\sum_\lambda
\int d^3k\left( a_{q\vec k\lambda }u_{\vec k\lambda }e^{i\vec k\cdot \vec x%
}+b_{q\vec k\lambda }^{\dagger }v_{\vec k\lambda }e^{-i\vec k\cdot \vec x%
}\right) .  \label{4}
\end{equation}

Insert eq.$\left( 4\right) $ into eq.$\left( 3\right) $ and after some
algebras, we obtain: 
\begin{eqnarray}
\vec \mu _q &=&\sum_\lambda \int d^3k\frac{Q_q}{2k_0}a_{q\vec k\lambda
}^{\dagger }i\vec \partial _k\times \vec ka_{q\vec k\lambda }  \nonumber \\
&&\ -\sum_\lambda \int d^3k\frac{Q_q}{2k_0}b_{q\vec k\lambda }^{\dagger }i%
\vec \partial _k\times \vec kb_{q\vec k\lambda }  \nonumber \\
&&\ +\sum_{\lambda \lambda' }\int d^3k\frac{Q_q}{2k_0}\chi _\lambda
^{\dagger }\vec \sigma \chi _{\lambda ^{\prime }}a_{q\vec k\lambda
}^{\dagger }a_{q\vec k\lambda ^{\prime }}  \nonumber \\
&&\ +\sum_{\lambda \lambda' }\int d^3k\frac{Q_q}{2k_0}\chi _\lambda
^{\dagger }\vec \sigma \chi _{\lambda ^{\prime }}b_{q\vec k\lambda ^{\prime
}}^{\dagger }b_{q\vec k\lambda }  \nonumber \\
&&\ -\sum_{\lambda \lambda' }\int d^3k\frac{Q_q}{2k_0}\chi _\lambda
^{\dagger }\frac{\vec \sigma \cdot \vec k}{2k_0\left( k_0+m_q\right) }i\vec 
\sigma \times \vec k\chi _{\lambda ^{\prime }}a_{q\vec k\lambda }^{\dagger
}a_{q\vec k\lambda ^{\prime }}  \nonumber \\
&&\ -\sum_{\lambda \lambda' }\int d^3k\frac{Q_q}{2k_0}\chi _\lambda
^{\dagger }\frac{\vec \sigma \cdot \vec k}{2k_0\left( k_0+m_q\right) }i\vec 
\sigma \times \vec k\chi _{\lambda ^{\prime }}b_{q\vec k\lambda ^{\prime
}}^{\dagger }b_{q\vec k\lambda }  \nonumber \\
&&\ -\sum_\lambda \int d^3k\frac{Q_q}{2k_0}\frac{\vec k}{%
2(k_0+m_q)}a_{q\vec k\lambda }^{\dagger }b_{q-\vec k\lambda 
}^{\dagger }+h.c.  \nonumber \\
&&\ +\sum_{\lambda \lambda ^{\prime }}\int d^3k\frac{Q_q}{2k_0}a_{q\vec k%
\lambda }^{\dagger }i\vec \partial _kb_{q-\vec k\lambda ^{\prime }}^{\dagger
}\times \chi _\lambda ^{\dagger }\left( m_q\vec \sigma +\frac{\vec \sigma
\cdot \vec k}{k_0+m_q}i\vec \sigma \times \vec k\right) \chi _{\lambda
^{\prime }}+h.c.  \label{5}
\end{eqnarray}

Now we see that the first two terms are almost the standard forms of orbital
magnetic moment in nonrelativistic quantum mechanics expressed in momentum
space, except that the static quark mass $m_q$ is now replaced by the
relativistic quark energy $k_0$. The third and fourth terms (where $\chi
_\lambda $ denotes the Pauli spinor) also resemble the expression of spin
magnetic moment in nonrelativistic quantum mechanics, with again the static
quark mass replaced by the relativistic quark energy. However, compared with
eq.(1), we have four more terms: the fifth and sixth terms are relativistic
corrections; and the last two terms are quark-antiquark pair creation and
annihilation terms which are characteristic of quantum field theory and only
contribute to baryon magnetic moments when sea quark excitation mixing is
included is baryon ground states.

Eq.(5) is not yet the final expression we aimed at, since the first four
terms correspond to the orbital and spin contributions in nonrelativistic
quantum mechanics, while in quantum chromodynamics, the quark spin and
orbital angular momentum operators are 
\begin{eqnarray}
\vec Sq &=&\frac 12\int d^3x\bar \psi _q\vec \gamma \gamma ^5\psi _q,
\label{6} \\
\vec Lq &=&\int d^3x\psi _q^{\dagger }\vec x\times \frac 1i\vec \partial
\psi _q.  \label{7}
\end{eqnarray}

Again insert eq.$\left( 4\right) $ into them and transit to the momentum
space, we get: 
\begin{eqnarray}
\vec Sq &=&\frac 12\sum_{\lambda \lambda' }\int d^3k\chi _\lambda
^{\dagger }\left( \vec \sigma -\frac{\vec \sigma \cdot \vec k}{k_0\left(
k_0+m_q\right) }i\vec \sigma \times \vec k\right) \chi _{\lambda ^{\prime
}}a_{q\vec k\lambda }^{\dagger }a_{q\vec k\lambda ^{\prime }}  \nonumber \\
&&\ -\frac 12\sum_{\lambda \lambda' }\int d^3k\chi _\lambda ^{\dagger
}\left( \vec \sigma -\frac{\vec \sigma \cdot \vec k}{k_0\left(
k_0+m_q\right) }i\vec \sigma \times \vec k\right) \chi _{\lambda ^{\prime
}}b_{q\vec k\lambda ^{\prime }}^{\dagger }b_{q\vec k\lambda }  \nonumber \\
&&\ +\frac 12\sum_{\lambda \lambda ^{\prime }}\int d^3k\chi _\lambda
^{\dagger }\frac{i\vec \sigma \times \vec k}{k_0}\chi _{\lambda ^{\prime
}}a_{q\vec k\lambda }^{\dagger }b_{q-\vec k\lambda ^{\prime }}^{\dagger
}+h.c.  \nonumber \\
\ &\equiv &\int d^3k\vec S_{q\vec k}+\int d^3k\vec S_{\bar q\vec k}+{\rm %
pair~creation~(annhination)~term}  \nonumber \\
\ &\equiv &\vec S_q+\vec S_{\bar q}+{\rm pair~creation~(annhination)~term},
\label{8}
\end{eqnarray}
\begin{eqnarray}
\vec Lq &=&\int d^3k\left( \sum_\lambda a_{q\vec k\lambda }^{\dagger }i\vec 
\partial _k\times \vec ka_{q\vec k\lambda }+\sum_{\lambda \lambda' }\chi
_\lambda ^{\dagger }\frac{\vec \sigma \cdot \vec k}{2k_0\left(
k_0+m_q\right) }i\vec \sigma \times \vec k\chi _{\lambda ^{\prime }}a_{q\vec 
k\lambda }^{\dagger }a_{q\vec k\lambda ^{\prime }}\right)  \nonumber \\
&&+\int d^3k\left( \sum_\lambda b_{q\vec k\lambda }^{\dagger }i\vec \partial
_k\times \vec kb_{q\vec k\lambda }-\sum_{\lambda \lambda' }\chi _\lambda
^{\dagger }\frac{\vec \sigma \cdot \vec k}{2k_0\left( k_0+m_q\right) }i\vec 
\sigma \times \vec k\chi _{\lambda ^{\prime }}b_{q\vec k\lambda ^{\prime
}}^{\dagger }b_{q\vec k\lambda }\right)  \nonumber \\
&&-\sum_{\lambda \lambda ^{\prime }}\int d^3k\chi _\lambda ^{\dagger }\frac{i%
\vec \sigma \times \vec k}{2k_0}\chi _{\lambda ^{\prime }}a_{q\vec k\lambda
}^{\dagger }b_{q-\vec k\lambda ^{\prime }}^{\dagger }+h.c.  \nonumber \\
&\equiv &\int d^3k\vec L_{q\vec k}+\int d^3k\vec L_{\bar q\vec k}+{\rm %
pair~terms}  \nonumber \\
&\equiv &\vec L_q+\vec L_{\bar q}+{\rm pair~terms}.  \label{9}
\end{eqnarray}

Now compare eqs.$\left( 5\right) $, $\left( 8\right) $ and $\left( 9\right) $%
, we find that the magnetic moment operator in relativistic quantum field
theory can be expressed in a quite elegant form which is exactly analogous
to our notion of magnetic moment in nonrelativistic quantum mechanics except
for the additional quark-antiquark pair creation (annihilation) term: 
\begin{equation}
\vec \mu _q=\int d^3k\frac{Q_q}{k_0}\left( \vec S_{q\vec k}-\vec S_{\bar q%
\vec k}\right) +\int d^3k\frac{Q_q}{2k_0}\left( \vec L_{q\vec k}-\vec L_{%
\bar q\vec k}\right) +{\rm pair~terms}.  \label{10}
\end{equation}

Applying this operator to baryon ground states as shown in eq.(2) and
approximating the $\frac 1{k_0}$ factors by the average value of
relativistic quark energy inside the baryon, we can express baryon magnetic
moment in the language of baryon spin structure: 
\begin{equation}
\mu _B=\sum_q\frac{Q_q}{2\left\langle k_0\right\rangle _q}\left( \Delta
_q-\Delta _{\bar q}+L_q-L_{\bar q}\right) +{\rm pair~terms},  \label{11}
\end{equation}
where $L_q$ ($L_{\bar q}$) denotes the quark (antiquark) orbital
contribution to the baryon spin, and $\Delta _q$ ($\Delta _{\bar q}$)
represents the proportion of baryon spin carried by quark (antiquark) spin
or helicity.

Eqs. $\left( 10\right) $ and $\left( 11\right) $ are elegant by their forms
and physical interpretations. However, $L_q-L_{\bar q}$ has less direct
experimental significance. In the following we will make use of the
measurable baryon tensor charge and rearrange eq.(5) into more practical
expressions.

Baryon tensor charge $\delta q$ is defined through 
\begin{equation}
\left\langle PS\left| \int d^3x\bar \psi _q\sigma ^{\mu \nu }\psi _q\right|
PS\right\rangle =\bar u_{PS}\sigma ^{\mu \nu }u_{PS}\cdot \delta q.
\label{12}
\end{equation}
It has received an increasing amount of attention after it was shown by
Jaffe and Ji that it can be related to the first moment of quark
transversity distribution $h_1\left( x\right) $ measured in transversely
polarized scattering experiments \cite{4}. For our purpose we consider a
baryon at rest and polarized along the third direction, in which case $%
\delta q=\left\langle PS\left| (\vec \delta q)_3\right| PS\right\rangle $,
with 
\begin{equation}
\vec \delta q=\int d^3x\bar \psi _q\vec \Sigma \psi _q.  \label{13}
\end{equation}
Insert eq.$\left( 4\right) $ into eq$.\left( 13\right) $, we get: 
\begin{eqnarray}
\vec \delta q &=&\sum_{\lambda \lambda' }\int d^3k\chi _\lambda ^{\dagger
}\left( \frac{m_q}{k_0}\vec \sigma +\frac{\vec \sigma \cdot \vec k}{%
k_0\left( k_0+m_q\right) }i\vec \sigma \times \vec k\right) \chi _{\lambda
^{\prime }}a_{q\vec k\lambda }^{\dagger }a_{q\vec k\lambda ^{\prime }} 
\nonumber \\
&&+\sum_{\lambda \lambda' }\int d^3k\chi _\lambda ^{\dagger }\left( \frac{%
m_q}{k_0}\vec \sigma +\frac{\vec \sigma \cdot \vec k}{k_0\left(
k_0+m_q\right) }i\vec \sigma \times \vec k\right) \chi _{\lambda ^{\prime
}}b_{q\vec k\lambda ^{\prime }}^{\dagger }b_{q\vec k\lambda }  \nonumber \\
&&-\sum_\lambda \int d^3k\frac{\vec k}{k_0}a_{q\vec k\lambda }^{\dagger
}b_{q-\vec k\lambda }^{\dagger }+h.c.  \nonumber \\
&\equiv &\vec \delta _q-\vec \delta _{\bar q}+{\rm pair~terms}.  \label{14}
\end{eqnarray}

Compare eqs. (5), (8), and (14), and also approximate $\frac 1{k_0}$ factors
by the average value of relativistic quark energy inside the baryon, we see
that baryon magnetic moment can be expressed as 
\begin{eqnarray}
\vec \mu &=&\sum_q\frac{Q_q}{2\left\langle k_0\right\rangle _q}\left( \vec L%
_q^{NR}-\vec L_{\bar q}^{NR}\right)  \nonumber \\
&&\ \ +\sum_q\frac{Q_q}{\left\langle k_0\right\rangle _q}\left( 1-\frac{m_q}{%
2\left( \left\langle k_0\right\rangle _q+m_q\right) }\right) \left( \vec S_q-%
\vec S_{\bar q}\right) +\sum_q\frac{Q_q}{2\left( \left\langle
k_0\right\rangle _q+m_q\right) }\frac 12\vec \delta q  \nonumber \\
&&\ \ +\sum_{q\lambda \lambda ^{\prime }}\int d^3k\frac{Q_q}{2k_0}a_{q\vec k%
\lambda }^{\dagger }i\vec \partial _kb_{q-\vec k\lambda ^{\prime }}^{\dagger
}\times \chi _\lambda ^{\dagger }\left( m_q\vec \sigma +\frac{\vec \sigma
\cdot \vec k}{k_0+m_q}i\vec \sigma \times \vec k\right) \chi _{\lambda
^{\prime }}+h.c.,  \label{15}
\end{eqnarray}
where $\vec L_q^{NR}$ and $\vec L_{\bar q}^{NR}$ correspond to the
derivative terms in the expression of $\vec Lq$ in eq.(9) and can be
recognized as ``nonrelativistic'' orbital angular momentum in
nonrelativistic quark models. For ground state baryons, it is plausible to
assume $L_q^{NR}(L_{\bar q}^{NR})=0$, as in most quark models. In contrast,
as can be seen from eq.(9), the ``relativistic'' orbital angular momentum $%
L_q$ and $L_{\bar q}$ will not be zero even for pure $s$-wave state. And the 
$\frac 12\vec \delta q$ term in eq.(15) is just the ``trace'' left by such
``relativistic'' orbital contribution.

The quark-antiquark pair creation (annihilation) terms only contribute when
different Fock components mixing is taken into account. A constituent quark
model with valence $q^3$ and sea $q^3q\bar q$ mixing has been used to
estimate the contribution of such terms to magnetic moment, and it is found
to be small ($-0.065\mu _P$ for proton) \cite{5}. With the assumptions: 
\begin{eqnarray}
&&L_q^{NR}=L_{\bar q}^{NR}=0  \nonumber \\
&&{\rm Fock~components~mixing~contributions~negligible,}  \label{16}
\end{eqnarray}
eq.(15) reduces to 
\begin{equation}
\vec \mu =\sum_q\frac{Q_q}{\left\langle k_0\right\rangle _q}\left[ \left( 1-%
\frac{m_q}{2\left( \left\langle k_0\right\rangle _q+m_q\right) }\right)
\left( \vec S_q-\vec S_{\bar q}\right) +\frac{\left\langle k_0\right\rangle
_q}{2\left( \left\langle k_0\right\rangle _q+m_q\right) }\frac 12\vec \delta
q\right] .  \label{17}
\end{equation}
Or equivalently, 
\begin{equation}
\mu _B=\sum_q\frac{Q_q}{2\left\langle k_0\right\rangle _q}\left[ \left( 1-%
\frac{m_q}{2\left( \left\langle k_0\right\rangle _q+m_q\right) }\right)
\left( \Delta _q-\Delta _{\bar q}\right) +\frac{\left\langle
k_0\right\rangle _q}{\left\langle k_0\right\rangle _q+m_q}\frac 12\delta
q\right] .  \label{18}
\end{equation}

This is a good expression to work with, for it expresses baryon magnetic
moment through baryon tensor charge and quark (antiquark) contribution to
baryon spin, which are all planned to be measured experimentally. By
assuming SU(3) flavor symmetry among the octet baryons, their magnetic
moments can be expressed in terms of the spin contents of the proton \cite
{2,3}: 
\begin{eqnarray}
\mu _P &=&\frac{2e}{6\left\langle k_0\right\rangle _u}W_u+\frac{-e}{%
6\left\langle k_0\right\rangle _d}W_d+\frac{-e}{6\left\langle
k_0\right\rangle _s}W_s,  \nonumber \\
\mu _n &=&\frac{-e}{6\left\langle k_0\right\rangle _d}W_u+\frac{2e}{%
6\left\langle k_0\right\rangle _u}W_d+\frac{-e}{6\left\langle
k_0\right\rangle _s}W_s,  \nonumber \\
\mu _{\Sigma ^{+}} &=&\frac{2e}{6\left\langle k_0\right\rangle _u}W_u+\frac{%
-e}{6\left\langle k_0\right\rangle _s}W_d+\frac{-e}{6\left\langle
k_0\right\rangle _d}W_s,  \nonumber \\
\mu _{\Sigma ^{-}} &=&\frac{-e}{6\left\langle k_0\right\rangle _d}W_u+\frac{%
-e}{6\left\langle k_0\right\rangle _s}W_d+\frac{2e}{6\left\langle
k_0\right\rangle _u}W_s,  \nonumber \\
\mu _{\Xi ^0} &=&\frac{-e}{6\left\langle k_0\right\rangle _s}W_u+\frac{2e}{%
6\left\langle k_0\right\rangle _u}W_d+\frac{-e}{6\left\langle
k_0\right\rangle _d}W_s,  \nonumber \\
\mu _{\Xi ^{-}} &=&\frac{-e}{6\left\langle k_0\right\rangle _s}W_u+\frac{-e}{%
6\left\langle k_0\right\rangle _d}W_d+\frac{2e}{6\left\langle
k_0\right\rangle _u}W_s,  \nonumber \\
\mu _\Lambda &=&\frac 16\left( \frac{2e}{6\left\langle k_0\right\rangle _u}+%
\frac{-e}{6\left\langle k_0\right\rangle _d}\right) \left(
W_u+4W_d+W_s\right)  \nonumber \\
&&\ \ \ \ +\frac 16\cdot \frac{-e}{6\left\langle k_0\right\rangle _s}\left(
4W_u-2W_d+4W_s\right) ,  \nonumber \\
\mu _{\Sigma \Lambda } &=&\frac{-1}{2\sqrt{6}}\left( \frac{2e}{6\left\langle
k_0\right\rangle _u}-\frac{-e}{6\left\langle k_0\right\rangle _d}\right)
\left( W_u-2W_d+W_s\right) .  \label{19}
\end{eqnarray}
where 
\begin{equation}
W_q=\left( 1-\frac{m_q}{2\left( \left\langle k_0\right\rangle _q+m_q\right) }%
\right) \left( \Delta _q-\Delta _{\bar q}\right) +\frac{\left\langle
k_0\right\rangle _q}{\left\langle k_0\right\rangle _q+m_q}\frac 12\delta q,
\label{20}
\end{equation}
and $\Delta _q$, $\Delta _{\bar q}$ etc. all refer to the quantities of the
proton.

With the measurements of $\Delta _q$, $\Delta _{\bar q}$ and $\delta q$, and
together with the existing experimental values of baryon magnetic moments,
eqs.(19) and (20) will provide a check of the assumptions in eq.(16) and
also of the SU(3) flavor symmetry.

Now lets go back to look at nonrelativistic constituent quark model. For the
first thing, as was noticed historically \cite{6}, and as we made explicitly
in eqs.(5), (10), (15) etc., 
it is not the static quark mass but the relativistic quark
energy inside the baryon that determines the quark contribution to baryon
magnetic moment. Due to confinement, quark kinetic energy inside the baryon
is rather high (an estimation with the uncertainty relation shows that it is
about 300MeV). This explains why constituent quark model can reproduce
baryon magnetic moment by adopting a large constituent quark mass. (There are 
also discussions on other mechanisms of producing a
large effective constituent quark mass, such as from chiral symmetry breaking
in chiral quark models \cite{7}.)

On the other hand, following the same steps as that lead to eq.(18), we can
obtain a parameterization of baryon magnetic moment similar to that in
nonrelativistic constituent quark model: 
\begin{eqnarray}
\mu _B &=&\sum_q\frac{Q_q}{2\left\langle k_0\right\rangle _q}\left[ \left( 1+%
\frac{m_q}{2\left\langle k_0\right\rangle _q}\right) \left( \Delta
_q^{NR}-\Delta _{\bar q}^{NR}\right) -\frac 12\left( \delta _q-\delta _{\bar 
q}\right) \right]   \label{21} \\
\  &=&\sum_q\frac{Q_q}{2\left\langle k_0\right\rangle _q}\left( 1+\frac{m_q}{%
2\left\langle k_0\right\rangle _q}-\frac 12\frac{\delta _q-\delta _{\bar q}}{%
\Delta _q^{NR}-\Delta _{\bar q}^{NR}}\right) \left( \Delta _q^{NR}-\Delta _{%
\bar q}^{NR}\right)   \label{22} \\
\  &\equiv &\sum_q\frac{Q_q}{2m_q^{eff}}\left( \Delta _q^{NR}-\Delta _{\bar q%
}^{NR}\right)   \label{23}
\end{eqnarray}
where 
\begin{eqnarray}
\Delta _q^{NR} &=&\left\langle PS\left| \sum_{\lambda \lambda ^{\prime
}}\int d^3k\chi _\lambda ^{\dagger }\sigma _3\chi _{\lambda ^{\prime }}a_{q%
\vec k\lambda }^{\dagger }a_{q\vec k\lambda ^{\prime }}\right|
PS\right\rangle ,  \nonumber \\
\Delta _{\bar q}^{NR} &=&\left\langle PS\left| -\sum_{\lambda \lambda
^{\prime }}\int d^3k\chi _\lambda ^{\dagger }\sigma _3\chi _{\lambda
^{\prime }}b_{q\vec k\lambda ^{\prime }}^{\dagger }b_{q\vec k\lambda
}\right| PS\right\rangle ,  \label{24}
\end{eqnarray}
correspond to the quark (antiquark) spin in nonrelativistic quark models,
and we have defined the factor times $\frac 12 Q_q \left( \Delta
_q^{NR}-\Delta _{\bar q}^{NR}\right) $ as the reciprocal of an ``effective''
constituent quark mass. In the absence of relativistic corrections and
antiquark polarization, $\left( \delta _q-\delta _{\bar q}\right) $ equals $%
\left( \Delta _q^{NR}-\Delta _{\bar q}^{NR}\right) $, and hence $m_q^{eff}$
is proportional to $\left\langle k_0\right\rangle _q$. Preliminary
experimental results show that antiquark polarization inside the nucleon is
rather small \cite{8}. And from an alternative expression of tensor charge 
\begin{equation}
\vec \delta q=\sum_{\lambda \lambda ^{\prime }}\int d^3k\chi _\lambda
^{\dagger }\left( \vec \sigma -\frac{(\vec \sigma \cdot \vec k)\vec k}{%
k_0\left( k_0+m\right) }\right) \chi _{\lambda ^{\prime }}\left( a_{\vec k%
\lambda }^{\dagger }a_{\vec k\lambda ^{^{\prime }}}+b_{\vec k\lambda
^{^{\prime }}}^{\dagger }b_{\vec k\lambda }\right) +{\rm pair~terms,}
\label{25}
\end{equation}
we see that the relativistic correction of tensor charge (the $(\vec \sigma
\cdot \vec k)\vec k$ term) is roughly of the order $k_3^2/k_0^2\approx 1/3$
for light quarks. Therefore we see from eqs. (22) and (23) that $m_q^{eff}$
is proportional to $\left\langle k_0\right\rangle _q$ within the error of
1/6. This finishes our analyses of why nonrelativistic constituent quark
models can explain baryon magnetic moment better than 20\%, by using the
simple expression of eq.(23) and adopting only two parameters of constituent
quark masses.

To end up our discussions, we must stress that through out this paper we
have not paid any attention to whether a baryon could be approximated as 
several constituent quarks at low energy. What we have provided for 
nonrelativistic constituent quark model is a justification of its simple 
magnetic-moment-spin relation as in eq.(23), which is the {\it essential} 
of its success at the magnetic moment problem. Eq.(23) will hold (within 
the error of 1/6 and with $m_q^{eff}$ roughly of the order of constituent 
quark mass) no matter how 
complicated the actual structure of the baryon is, and no matter    
whether there is really the picture of constituent quarks inside the 
baryon at all. 

We thank Bo-Qiang Ma and Hanxin He for helpful discussions. This work is
supported in part by NSFC(19675018), SEDC and SSTC of China.

\end{document}